\documentclass{article}
\usepackage{spconf,amsmath,graphicx,multicol,multirow,color,booktabs}
\usepackage[numbers,sort&compress]{natbib}
\title{A Novel Residual-guided Learning Method for Image Steganography}

\name{Miaoxin Ye, Dongxia Huang, Kangkang Wei, Weiqi Luo*\thanks{*W. Luo is the corresponding author.}}
\address{GuangDong Province Key Laboratory of Information Security Technology, \\  {School of Computer Science and Engineering, Sun Yat-sen University, GuangZhou, China}}
 
\begin{document}
\maketitle
\begin{abstract}
Traditional steganographic techniques have often relied on manually crafted attributes related to image residuals. These methods demand a significant level of expertise and face challenges in integrating diverse image residual characteristics. 
In this paper, we introduce an innovative deep learning-based methodology that seamlessly integrates image residuals, residual distances, and  image local variance to autonomously learn  embedding probabilities. 
Our framework includes an embedding probability generator and three pivotal guiding components: Residual guidance strives to facilitate embedding in complex-textured areas. Residual distance guidance aims to minimize the residual differences between cover and stego images. Local variance guidance effectively safeguards against modifications in regions characterized by uncomplicated or uniform textures. The three components collectively guide the learning process, enhancing the security performance.
 Comprehensive experimental findings underscore the superiority of our approach when compared to traditional steganographic methods and randomly initialized ReLOAD in the spatial domain.  

\end{abstract}

\section{Introduction}
\label{sec:intro}
Image steganography is the art of concealing secret information within an image for covert communication. Current steganography research is primarily based on the distortion minimization framework \cite{distortionMinimization}. Techniques such as Syndrome Trellis Coding (STC) \cite{stc} are subsequently used to approach the theoretical payload-distortion bound for practical embedding.   Traditional steganographic methods have often heuristically determined embedding costs, typically by analyzing on image residuals. {For instance}, HUGO \cite{hugo} measures embedding distortion based on the disparity between modified  SPAM \cite{spam} features and original cover image features. 
WOW \cite{wow} and S-UNIWARD \cite{uniward} both utilize residuals extracted by multi-directional wavelet filters for distortion calculation. 
Similarly, HILL \cite{hill} makes use of residuals but integrates modifications through low-pass filters, leading to enhanced security. 
However, what's noteworthy is the absence of a combined approach, leveraging both image residuals itself and relevant  statistical characteristics  (such as the texture difference between cover and stego), to augment steganographic security. Moreover, the dependency on handcrafted features in these methods necessitates a deep level of expertise. 

Recent advancements in steganography through deep learning can be broadly categorized into two main groups: those  based on Generative Adversarial Networks (GANs) \cite{ASDLGAN, UTGAN, spar}, and those based on adversarial examples \cite{advEmb, minmaxTIFS, enSGS}. Despite their accomplishments in bolstering security, these techniques exhibit certain limitations. For instance, existing methods necessitate the incorporation of a steganalytic discriminator to update the embedding  probabilities or costs. Consequently, GAN-based approaches grapple with issues such as training instability and mode collapse, arising primarily from the imbalance between the generator and discriminator. In parallel, methods relying on adversarial examples usually demand pre-training on established steganographic techniques, effectively refining existing methods.
Different from previous methodologies, ReLOAD \cite{ReLOAD} stands out as the pioneer steganography technique that operates without a discriminator. With a reinforcement learning strategy, it fomulates the reward soley with the residual distance between cover and stego images to modulate asymmetric costs.  However, it still relies significantly on existing methods like HILL and  MiPOD \cite{mipod}.  As a result, it faces limitations in terms of learning optimal costs from scratch.

 \begin{figure*}[t]
\centering
\includegraphics[scale=0.39]{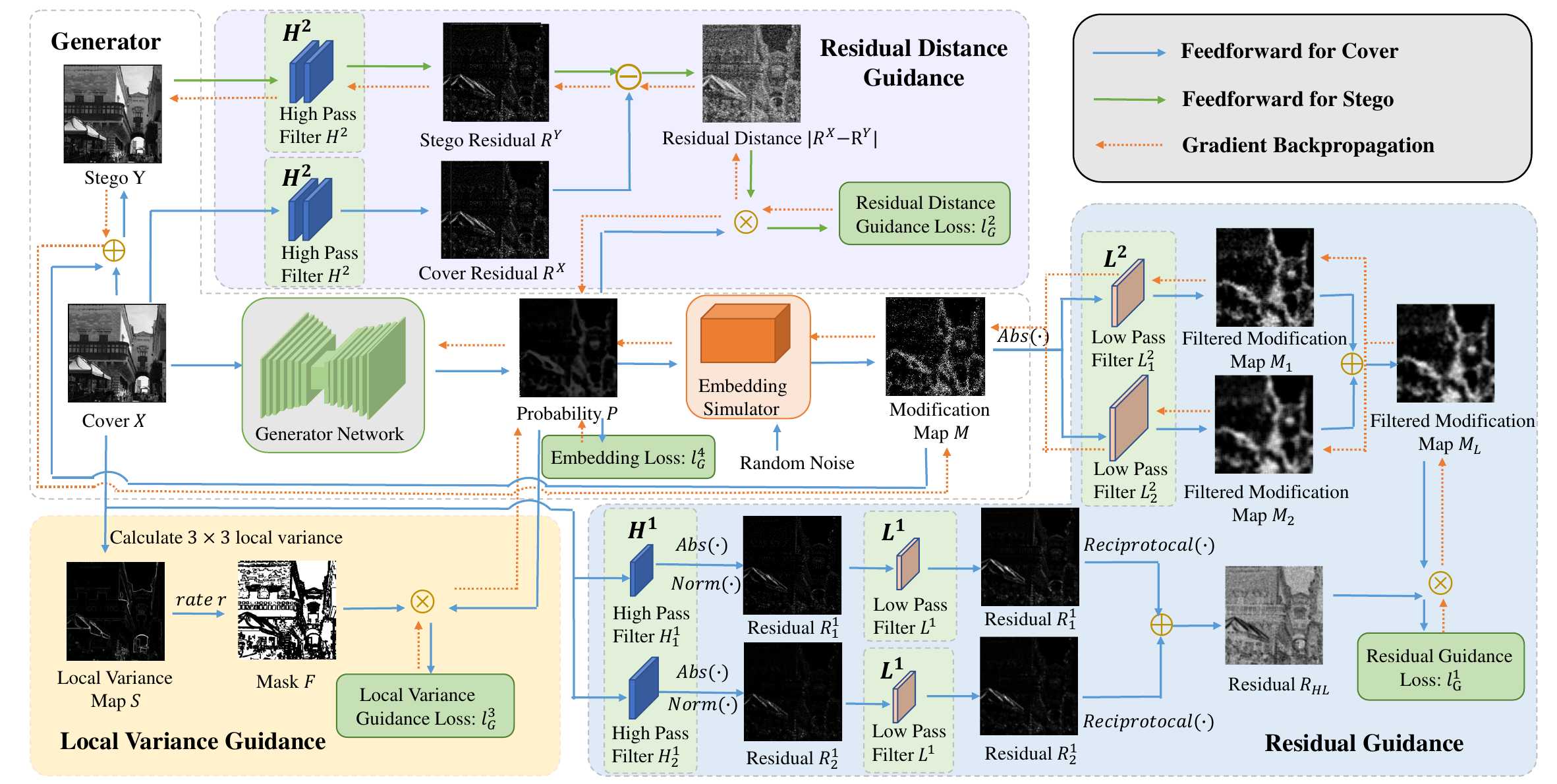} 
\caption{The structure of our model includes a generator and three guidance components, without the inclusion of a discriminator.} 
\label{fig:framework}
\end{figure*}

 To address above limitations, this paper introduces an innovative deep learning-based approach. By fusing various image residual characteristics, the proposed method introduces three guidance components to  cooperatively facilitate the embedding probability learning process: residual guidance steers embedding towards regions with complex textures, residual distance guidance minimizes texture discrepancies between cover and stego images, and local variance guidance discourages embedding within simple-textured areas identified by local variance. The proposed method, although employing a network, stands in contrast to existing deep learning-based methods. It not only dispenses with the need for a discriminator but also operates independently of existing steganographic techniques. Our method demonstrates that proper guidance, rather than a discriminator or delicate initialization, can also empower the generator network to learn embedding probabilities and produce secure stego images, which shed light on future deep learning-based methods. 
The main contributions are summarized as follows:  1) Compared to traditional steganography, our method adeptly combines diverse image residual attributes to collectively inform automatic embedding probability learning; 2) In contrast to established deep learning-based techniques, our approach sidesteps the reliance on a discriminator and severs ties with existing steganography methods; 3) Empirical findings on the BOSSBase \cite{BOSSBase} dataset underscore the superior security performance of our method relative to traditional approaches and ReLOAD without the use of existing steganography.

\section{Proposed Method}
\label{sec:method}
As shown in  Fig. \ref{fig:framework}, our model includes a generator and three  guidance components.  The generator initially employs a U-Net architecture to transform the cover  $X$ into a  probability map $P$. Subsequently, we simulate STC by employing the double-tanh activation function to generate the modification map $M$. Ultimately, the stego image is produced by $Y = X + M$.   Within this proposed framework, the application of these three distinct guidance components serves to constrain either $P$ or $M$ in the generator, a concept that will be expounded upon in the subsequent subsections.  Finally, we delve into the discussion of our loss functions used in our model. 

\subsection{Residual Guidance}
\label{ssec:residual guidance}
The residual guidance aims to steer embedding towards complex-textured regions with higher absolute values of cover image residuals.
As depicted in the bottom-right corner of Fig. \ref{fig:framework}, the residual guidance consists of three groups: the high-pass group $H^1$ and two low-pass groups $L^1$ and $L^2$. The high-pass group $H^1$ extracts the residuals of the cover $X$ with two high-pass filters KB3 and KV5. The absolute values of them are normalized and then filtered by the low-pass group $L^1$ to spread the values to their neighborhood, utilizing a $3 \times 3$ mean filter. The filtered residual maps are then merged by summing up their reciprocals and get $R_{HL}$. Meanwhile, the modification map $M$ with the absolute values is filtered by the low-pass group $L^2$ comprising two mean filters with sizes $11 \times 11$ and $7 \times 7$, to concentrate modification according to the diffusion principle. Then, the mean values of the filtered modification maps are calculated to get map $M_L$, which captures the information of both large and small filtering scales.
The residual guidance loss is eventually defined as follows: 
\begin{equation}
    l_G^1 = \frac{\|M_{L} \bigotimes R_{HL}\|_1}{h \times w},
\end{equation}
where $h$ and $w$ are the height and width of the cover $X$ and $\bigotimes$ is element-wise multiplication. In this way, $\|M_{L} \bigotimes R_{HL}\|_1$ denotes the sum of reciprocal values of residuals within the modified areas. Since our objective is to promote embedding in areas with large residual values, $\|M_{L} \bigotimes R_{HL}\|_1$ should be minimal, achieved by minimizing $l_G^1$.

\subsection{Residual Distance Guidance}
\label{ssec:residual distance guidance}
The  residual distance guidance primarily aims to minimize the residual gap between  cover and  stego, which has been identified as a way to enhance steganographic security in our prior study \cite{spp}. 
As depicted in the top-left corner of Fig. \ref{fig:framework}, this guidance firstly extracts the residual maps $R^X$ and $R^Y$ of  cover $X$ and  stego $Y$ produced by the generator with the high-pass group $H^2$, which comprises 30 fixed SRM \cite{SRM} filters. And then, the residual distance guidance loss is as follows:
\begin{equation}
\label{equ:rdg}
    l_G^2 = \frac{\| \frac{1}{30}\sum\limits_{k=1}^{30}|R^X_k-R^Y_k| \bigotimes P\|_1}{h \times w},
\end{equation}
where $R^X_k$ and $R^Y_k$ represents the $k$-th channel in $R^X$ and $R^Y$. As illustrate in (\ref{equ:rdg}), we additionally impose a constraint on $P$ to decrease the embedding probabilities within the residual domain, further reducing the residual distance. 

\subsection{Local Variance Guidance}
\label{ssec:complex region guidance}
Local variance serves as a metric for assessing image regions at a local level, frequently associating higher local variance with heightened residual values. Consequently, local variance can be considered an indicative trait of residuals.  The purpose of the local variance guidance is to discourage embedding in regions marked by low local variance, which are prone to easy detection by steganalyzers.

As depicted in the bottom-left corner of Fig. \ref{fig:framework}, the local variance is computed for the neighboring $3\times3$ pixels in cover $X$, leading to a local variance map $S$. To filter out pixels located in smoother or flat regions, a threshold value $r\in (0,1)$ is employed to identify regions characterized by the lowest local variances. This procedure results in a binary mask $F$, where 0 designates areas with relatively intricate textures in cover, while 1 indicates smooth regions. The local variance guidance loss is defined as follows:
\begin{equation}
    l_G^3 = \frac{\|F \bigotimes P\|_1}{h \times w}.
\end{equation}
To prevent embedding in smooth regions, it is crucial to minimize the embedding probabilities in these areas. Here, $\|F \bigotimes P\|_1$ is the sum of embedding probabilities within the smooth regions identified by mask $F$. The achievement of minimal values in it is facilitated by the reduction of $l_G^3$.  

\begin{table}[t]\small
    \renewcommand\arraystretch{1.1}
	\centering
	\caption{Detection error rate (\%) of our method with different $H^1$. Note that  values with an asterisk(*) denote the best results in the corresponding cases in all tables in this paper.}
   \vspace{0.1em}
 	\label{tab:H1}
	\begin{tabular}{ c c c c c c}
		\hline
		$H^1$ & KB3  &  KV5 &SRM6 & SRM30 & KB3+KV5  \\\hline 
		SRM  & 26.50  & 26.68  & 26.78 & 26.23 & \textbf{27.49*} \\   
        CovNet   & 17.98  & 18.33 & 18.22 & 17.78 & \textbf{19.25*}  \\\hline
	\end{tabular}
\vspace{-1em}
\end{table}

\subsection{Loss Function} 
In addition to the loss functions defined in the aforementioned three guidelines, we introduce the following embedding loss $l_G^4$  to ensure the integrity of the embedded payload: 
\begin{equation}
    l_G^4 = (-\sum\limits_{\vee (i,j)}\sum\limits_{\vee m} p^{(m)}_{i,j} \cdot \log_2 p^{(m)}_{i,j} - h \times w \times q )^2.
\end{equation}
Here, $m \in \{-1,0,1\}$ and $q$ represents the payload.
Finally, the total generator loss $l_G$ is defined as follows:
\begin{equation}
    l_G = \alpha \cdot l_G^1 + \beta \cdot l_G^2 + \gamma \cdot l_G^3 + \delta \cdot l_G^4.
\end{equation}
In this equation, we set $\alpha = 1$, $\beta = 10$, $\gamma = 1000$, and $\delta = 1 \times 10^{-4}$ to maintain a reasonable modification rate during the training process and to balance the magnitudes of the various loss components, with a larger magnitude for the more influential one. 

\section{Experimental Results}
\label{sec:experiment}

In our experiments, we firstly trained the proposed model using 40,000 images from SZUBase \cite{ASDLGAN} and subsequently assessed its security performance on 10,000 images from BOSSBase \cite{BOSSBase}. During the testing phase, we randomly selected 4,000 cover-stego pairs for training, set aside 1,000 pairs for validation, and used the remaining 5,000 pairs for evaluation.  In line with practices from \cite{UTGAN}, all images were resized to dimensions of $256 \times 256$ via the MATLAB's imresize function. For a comprehensive comparison, we incorporated five steganographic methods: WOW, S-UNIWARD, MiPOD, HILL and ReLOAD at 0.2 and 0.4 bpp. Security was evaluated using four steganalyzers: SRM \cite{SRM}, Yedroudj-Net \cite{yedroudjnet}, SRNet \cite{srnet} and CovNet \cite{deng}. To ensure reproducibility,  we have provided the source code for our model online \footnote{Our souce codes are available at: https://github.com/YMXxtc/Deepsteg}.

\begin{table}[t]\small
    \renewcommand\arraystretch{1.1}
	\centering
	\caption{Detection error rate (\%) of our method with different $H^2$.}
    \vspace{0.1em}
	\label{tab:H2}
	\begin{tabular}{ c c c c c c}
		\hline
		$H^2$ & KB3  &  KV5 & KB3+KV5 &SRM6 & SRM30  \\\hline 
		SRM  & 27.23  & 27.28  & 26.84& 27.27 & \textbf{27.49*} \\   
        CovNet   & 18.82  & 19.12 & 19.19 & 19.02 & \textbf{19.25*}  \\\hline
	\end{tabular}
\vspace{-1em}
\end{table}

\begin{table*}[t!]\small
    \renewcommand\arraystretch{1.1}
	\centering
	\caption{ Detection error rate (\%) evaluated on four steganalyzers on BOSSBase.}
	\label{tab:security}
	\begin{tabular}{c c c c c c c c c}
		\hline
		Steganalyzer   & \multicolumn{2}{c}{SRM}  & \multicolumn{2}{c}{Yedroudj-Net} & \multicolumn{2}{c}{SRNet} &  \multicolumn{2}{c}{CovNet} \\
        \cmidrule(r){2-3} \cmidrule(r){4-5} \cmidrule(r){6-7} \cmidrule(r){8-9}
		Payload  & 0.2 bpp & 0.4 bpp  & 0.2 bpp & 0.4 bpp & 0.2 bpp   & 0.4 bpp & 0.2 bpp   & 0.4 bpp\\\hline
		WOW \cite{wow}   &  32.82 &  22.15   & 25.24  & 13.92  &  18.00 &  11.90 & 18.15  & 10.31\\\hline    
        S-UNIWARD \cite{uniward}   & 34.36  & 21.78    &  31.35 & 17.52  & 21.95  & 12.92&  22.06 & 12.22\\\hline 
        MiPOD \cite{mipod}   & 36.97  & 25.94    &  34.25 &  21.49 & 29.27  & 19.02 & 28.80  & 17.16 \\\hline
        HILL \cite{hill}   &  37.02 &  26.92   & 31.33  & 19.54  & 27.21  &  17.33 & 26.28  & 16.10 \\\hline
         ReLOAD-Rand \cite{ReLOAD}   & 10.00  & 7.60   & 6.20  & 4.45  & 4.90  &3.93 &  5.40 &  3.53\\\hline
        Proposed & \textbf{38.17*}  & \textbf{27.49*}    & \textbf{34.90*}  &  \textbf{23.39} &  \textbf{29.70*} & \textbf{20.18*}& \textbf{29.36*} & \textbf{19.25*}\\\hline  
	\end{tabular}
 \vspace{-1em}
\end{table*}

\subsection{Selection of Hyperparameters}
\label{ssec:hyperparameters}
Several crucial hyperparameters demand attention in our model, specifically the filters $H^1$, $H^2$ and the threshold $r$. In the following, we will delve deeper into the rationale behind our selections of them. For simplicity, all experiments were carried out at  the embedding payload of 0.4 bpp.  

\noindent \textbf{Selection of $H^1$.}  Several popular filters, including KB3, KV5,  SRM6 \cite{UTGAN}, SRM30 \cite{SRM}, and 'KB3+KV5' are included for comparative study. The results presented in Table \ref{tab:H1} highlight that the combination of KB3 and KV5 within $H^1$ yields the optimal security performance. Therefore we utilize this combination in the proposed method as depicted in Fig. \ref{fig:framework}.

\begin{table}[t]\small
    \renewcommand\arraystretch{1.1}
	\centering
	\caption{Detection error rate (\%) of our method with different $r$. The underlined values indicate the second-best outcomes.}
    \vspace{0.1em}
	\label{tab:rate}
	\begin{tabular}{ c c c c c c}
		\hline
		$r$ & 0.3  &  0.4 & 0.5 & 0.6 & 0.7  \\\hline 
		SRM  & \textbf{27.84*}  & 26.99 & \underline{27.49} & 26.79 & 26.09  \\   
        CovNet   & 18.75  & \underline{19.00} & \textbf{19.25*} & 18.64 & 17.53 \\\hline    
	\end{tabular}
 \vspace{-1em}
\end{table}

\noindent \textbf{Selection of $H^2$.}  $H^2$ is significant in the residual distance guidance part. We conduct tests on the same filters as $H^1$. As deomonstrated in Table \ref{tab:H2}, SRM30 achieves the best performance, which is utilized in the proposed method.

\noindent \textbf{Selection of $r$.}  The parameter $r$ governs the extent of the smooth region in local variance guidance. So we investigate the impact of rate $r$ ranging from 0.3 to 0.7. The results are shown in Table \ref{tab:rate}. Given that $r=0.5$ yields the optimal result against CovNet and a slightly inferior result against SRM, we adopt a rate of 0.5 for both 0.2 bpp and 0.4 bpp.

\subsection{Security Performance}
\label{ssec:security}
In this section, we conduct a security comparison of our method with five other related approaches. It's worth noting that our approach does not rely on any existing steganography or a discriminator. Therefore, we exclude methods that are based on GANs or adversarial examples. Besides, to ensure a fair and unbiased comparison, we introduce a randomly initialized cost variant of ReLOAD, denoted as ReLOAD-Rand.  

The results are shown in Table \ref{tab:security}. From Table \ref{tab:security}, we have two observations: 1) Our method outperforms traditional methods at both 0.2 bpp and 0.4 bpp, especially for CNN-based steganalyzers at 0.4 bpp. Taking CovNet at 0.4 bpp as an example, the detection error rate of the proposed method is 19.25\%, achieving around 8.94\%, 7.03\%, 2.09\% and 3.15\% achievement respectively compared with WOW, S-UNIWARD, MiPOD and HILL. 2) With random initialization, the detection error rates for ReLOAD-Rand are \textbf{all below 10\%}. The result indicates that ReLOAD is highly dependent on existing steganography.  By contrast, our method is independent of any existing method, enabling to learn embedding probability from scratch.

\begin{table}[t]\small 
    \renewcommand\arraystretch{1.1}
	\centering
	\caption{{Detection error rate (\%) of our method with different combinations of the three guidance components. }}
	\label{tab:abalation}
	\begin{tabular}{c c c c c}
		\hline
		Steganalyzer   &  SRM   & Yedroudj-Net & SRNet &  CovNet\\\hline
        RDG   &  6.68  & 4.10 & 3.81 &  3.34\\\hline
        LVG   & 19.87   & 12.29 & 11.95 &  10.59\\\hline
        RG   & 26.96   & 23.20 & 18.41 &  17.79\\\hline
        RDG+LVG   & 19.37   & 11.94 & 11.60 &  10.40\\\hline 
        RG+RDG   & 27.11   & 22.99 & 19.40 & 18.13\\\hline
        RG+LVG   & 27.43   & 23.11 & 20.02 &  19.11\\\hline
		RG+RDG+LVG  &  \textbf{27.49*}  & \textbf{23.39*} & \textbf{20.18*}&  \textbf{19.25*} \\\hline        
	\end{tabular}
\vspace{-1em}
\end{table}

\subsection{Ablation Study}
\label{ssec:ablation}
In this section, we assess the influence of the three guidance components incorporated into our model. To achieve this, we compare the performance of the proposed model with various combinations of these components. The results are presented in Table \ref{tab:abalation}, where 'RG,' 'RDG,' and 'LVG' denote residual guidance, residual distance guidance, and local variance guidance, respectively. Notably, our findings reveal that when a single guidance component is employed, residual guidance exhibits the highest level of security performance, followed by local variance guidance, with residual distance guidance performing the least effectively. The culmination of all three components in 'RG+RDG+LVG' results in the highest level of security performance, underscoring the synergistic impact of each component in enhancing the overall effectiveness, which is ultimately integrated into the proposed method.

\section{Conclusion}
\label{sec:conclusion}

This paper introduces a novel deep learning-based steganographic model guided by residuals. The proposed model exclusively exploits a wide range of image residual characteristics, akin to traditional steganographic methods. Notably, it operates independently of any pre-existing steganography techniques or a discriminator, which are commonly utilized in most existing deep learning-based approaches. Extensive comparative results demonstrate that the proposed method outperforms traditional methods and ReLOAD-Rand.

While this paper represents the initial endeavor to develop a steganographic model without the need for pre-existing steganography methods and a discriminator to learn embedding probabilities, there are still certain aspects worthy of further investigation. In addition to residual characteristics, we will also explore high-dimensional steganalytic features and consider asymmetric strategies, as discussed in \cite{Synch}. Furthermore, it's essential to highlight that our model has the capability to seamlessly integrate into modern steganographic frameworks based on adversarial samples or GANs, enabling continuous improvement of its performance.

\clearpage


\bibliographystyle{IEEEbib}
\bibliography{mybibfile}
\end{document}